# Rejection of randomly coinciding 2ν2β events in ZnMoO$_4$ scintillating bolometers[a]

D.M. Chernyak[1,2,b], F.A. Danevich[1], A. Giuliani[2], M. Mancuso[2,3], C. Nones[4], E. Olivieri[2], M. Tenconi[2] and V.I. Tretyak[1]

[1]Institute for Nuclear Research, MSP 03680 Kyiv, Ukraine
[2]Centre de Sciences Nucléaires et de Sciences de la Matière, 91405 Orsay, France
[3]Dipartimento di Scienza e Alta Tecnologia dell'Università dell'Insubria, I-22100 Como, Italy
[4]Service de Physique des Particules, CEA-Saclay, F-91191 Gif sur Yvette, France

**Abstract.** Random coincidence of 2ν2β decay events could be one of the main sources of background for 0ν2β decay in cryogenic bolometers due to their poor time resolution. Pulse-shape discrimination by using front edge analysis, the mean-time and $\chi^2$ methods was applied to discriminate randomly coinciding 2ν2β events in ZnMoO$_4$ cryogenic scintillating bolometers. The background can be effectively rejected on the level of 99% by the mean-time analysis of heat signals with the rise time about 14 ms and the signal-to-noise ratio 900, and on the level of 98% for the light signals with 3 ms rise time and signal-to-noise ratio of 30 (under a requirement to detect 95% of single events). Importance of the signal-to-noise ratio, correct finding of the signal start and choice of an appropriate sampling frequency are discussed.

## 1 Introduction

Observation of neutrinoless double beta (0ν2β) decay would imply the violation of lepton number conservation and definitely new physics beyond the Standard Model, establishing the Majorana nature of neutrino [1, 2, 3, 4, 5]. Cryogenic scintillating bolometers look the most promising detectors to search for this extremely rare process in a few theoretically favourable nuclei [6, 7, 8, 9, 10, 11, 12]. Zinc molybdate (ZnMoO$_4$) is one of the most promising materials to search for 0ν2β decay thanks to the absence of long-lived radioactive isotopes of constituting elements, comparatively high percentage of the element of interest and recently developed technique to grow large high quality radiopure ZnMoO$_4$ crystal scintillators [10, 11, 13, 14, 15, 16]. However, a disadvantage of the low temperature bolometers is their poor time resolution, which can lead to a significant background at the energy $Q_{2\beta}$ due to random coincidences of signals, in particular due to the unavoidable two-neutrino 2β decay events [17]. This is a notable problem for the experiments aiming to search for 0ν2β decay of $^{100}$Mo, because of a comparatively low half-life of $^{100}$Mo relatively to the two neutrino double beta decay $T_{1/2}$ = 7.1 × 10$^{18}$ yr [18]. Counting rate of two randomly coincident 2ν2β events in cryogenic Zn$^{100}$MoO$_4$ detectors is expected to be on the level of 2.9 × 10$^{-4}$ counts / (keV × kg × yr) at the $Q_{2\beta}$ energy (for 100 cm$^3$ crystals), meaning that randomly coincident 2ν2β decays can be even a main source of background in a future large scale high purity experiment [17].

## 2 Generation of randomly coinciding signals

Sets of single and randomly coincident signals were generated by using pulse profiles and noise baselines accumulated with 0.3 kg ZnMoO$_4$ crystal scintillator operated as a cryogenic scintillating bolometer in Centre de Sciences Nucléaires et de Sciences de la Matière (Orsay, France) with the sampling rate 5 kSPS both for the light and heat channels, and in the Modane underground laboratory (France) with the sampling rate 1.9841 kSPS for the both channels. Ten thousand of base-line samples were selected in all the cases.

The pulse profiles of heat and light signals of the detectors (sum of a few hundred pulse samples produced mainly by cosmic rays with energy of a few MeV) were obtained by fit with the following function:

$$f_S(t) = A \cdot \left(e^{-t/\tau_1} + e^{-t/\tau_2} - e^{-t/\tau_3} - e^{-t/\tau_4}\right), \quad (1)$$

where $A$ is the amplitude, $\tau_1$, $\tau_2$, $\tau_3$ and $\tau_4$ are the time constants.

---

To generate randomly coinciding signals in the region of $Q_{2\beta}$ value of $^{100}$Mo, the amplitude of the first pulse $A_1$ was obtained by sampling the $2\nu2\beta$ distribution for $^{100}$Mo, while the amplitude of the second pulse was chosen as $A_2 = Q_{2\beta} - A_1 + \Delta E$, where $\Delta E$ is a random component in the energy interval $[-5, +5]$ keV (which is a typical energy resolution of a bolometer).

Ten thousand coinciding signals were randomly generated in the time interval from 0 to $3.3 \cdot \tau_R$ ($\Delta t = 3.3 \cdot \tau_R$), where $\tau_R$ is the rise time of the signals (defined here as the time to change the pulse amplitude from 10% to 90% of its maximum). As it will be demonstrated in the next section, the rejection efficiency of randomly coinciding signals (RE, defined as a part of the pile-up events rejected by pulse-shape discrimination) reaches almost its maximal value when the time interval of consideration exceeds $(3 - 4) \tau_R$. Ten thousand of single signals were also generated.

A signal-to-noise ratio (defined as ratio of the maximum signal amplitude to the standard deviation of the noise baseline) was taken 30 for light signals and 900 for heat signals. These values are typical for $ZnMoO_4$ scintillating bolometers. Examples of the generated heat and light single pulses are presented in Fig. 1.

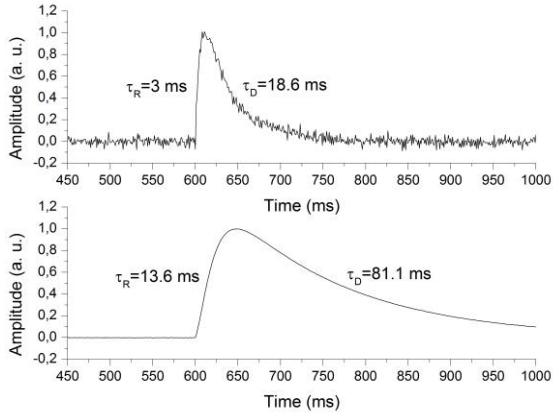

**Figure 1.** Examples of generated light (upper part) and heat (lower part) pulses. $\tau_R$ and $\tau_D$ denote rise and decay times, respectively.

## 3 Methods of pulse-shape discrimination

We have applied three methods to discriminate randomly coincident events: mean-time method, $\chi^2$ approach, and front edge analysis. We demanded a 95% efficiency in accepting single signals.

### 3.1 Mean-time method

The following formula was applied to calculate the parameter $\langle t \rangle$ (mean-time) for each pulse $f(t_k)$:

$$\langle t \rangle = \sum f(t_k) t_k / \sum f(t_k), \qquad (2)$$

where the sum is over time channels $k$, starting from the origin of a pulse and up to a certain time.

As a first step we have chosen the time interval $\Delta t$ to analyze efficiency of the pulse-shape discrimination. Six sets of single and randomly coinciding light (with $\tau_R = 3$ ms) and heat signals (with $\tau_R = 13.6$ ms) were generated in the time intervals from 1 to about 6 pulse rise times. The results of this analysis are presented in Fig. 2. The uncertainties of the rejection efficiency were estimated by analysis of three sets of data generated using three sets of different noise baseline profiles (about 3300 profiles in the each set). One can see that the rejection efficiency of randomly coinciding signals reaches its maximal value when the time interval $\Delta t$ is larger than $(3 - 4) \tau_R$. All the further analysis was done by using data generated in the time interval $\Delta t = 3.3 \tau_R$.

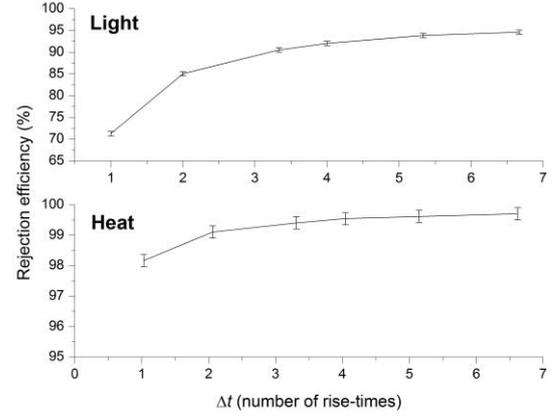

**Figure 2.** Dependence of the rejection efficiency (by using the mean-time method) for heat and light channels on the time interval $\Delta t$ where the randomly coinciding signals were generated.

Typical distributions of the mean time parameters for single and pile-up events are presented in Fig. 3. The rejection efficiency of randomly coinciding pulses, under the requirement to detect 95% of single events, is 90.5%.

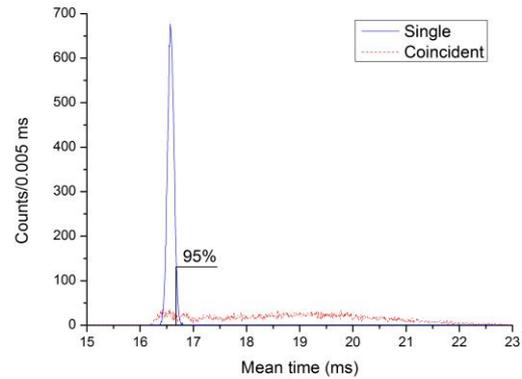

**Figure 3.** Distribution of the mean-time parameter for single and coincident light pulses with a rise time 3 ms. Rejection efficiency of coincident pulses is 90.5%. The events left from arrow are accepted as single events. 9.5% of pile-up events moves to the "single" events region due to incorrect start finding and / or too small difference between coinciding signals.

One could expect that rejection efficiency of pulse-shape discrimination depends on choice of the time interval used to calculate a discrimination parameter. For instance, in Fig. 4 the results of the mean-time method optimization are presented. Rejection efficiency has maximum when the mean-time parameter is calculated from the signal origin to 30th channel what roughly corresponds to ~ $\tau_D$. All the discrimination methods were optimized in a similar way.

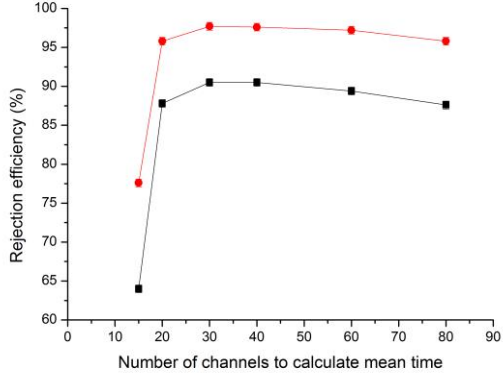

**Figure 4.** Dependence of the rejection efficiency of the mean-time method on the number of channels to calculate the parameter $\langle t \rangle$. The analysis was performed for the light signals with 3 ms rise time. Rejection efficiency of randomly coinciding pulses is 90.5% for the cases when the start of the signals was found by our algorithm (squares), and 97.7% for the known start position (circles). One channel is ≈ 0.504 ms.

The dependences presented in Fig. 4 demonstrate also importance to optimize an algorithm for signals start finding, particularly for the light pulses with comparatively low signal-to-noise ratio. Rejection efficiency is substantially higher in case when the start position of each pulse is known from the generation algorithm.

### 3.2 $\chi^2$ method

The approach is based on the calculation of the $\chi^2$ parameter defined as

$$\chi^2 = \sum (f(t_k) - f_S(t_k))^2 \qquad (3)$$

where the sum is over time channels $k$, starting from the origin of pulse and up to a certain time $t_n$. The number of channels to calculate the $\chi^2$ have been optimized to reach a maximal rejection efficiency.

### 3.3 Front edge analysis

The front edge parameter can be defined as the time between two points on the pulse front edge with amplitudes $Y_1$% and $Y_2$% of the pulse amplitude. The parameters $Y_1$ and $Y_2$ should be optimized to provide maximal rejection efficiency. For instance, the highest rejection efficiency for heat pulses was reached with the front edge parameter determined as time between the signal origin and the time where the signal amplitude is $Y_2 = 90$% of its maximum (RE = 98.7%).

However, the rejection efficiency of the front edge method is limited due to the some fraction of randomly coinciding events with a small first (with the amplitude $A_1$ below $Y_1$) or second pulse (with a low amplitude, and appearing well after the first signal maximum).

## 4 Results and discussion

The methods of pulse-shape discrimination are compared in Table 1. The data were obtained with start positions of signals known *a priory* from the generation procedure to avoid a substantial dependence on the signal start finding algorithm (it should be stressed that the accuracy of the algorithms depends on the sampling rate of the data acquisition, see below). Such an approach gives an upper limit of the methods efficiency. All the methods give 92% − 98% rejection efficiency by using light signal with the rise time 3 ms and ≈99% for much slower heat signals with the rise time 13.6 ms. One can conclude that the signal-to-noise ratio (30 for the light and 900 for the heat signals) plays a crucial role in the pulse-shape discrimination of randomly coinciding events in cryogenic bolometers. Analysis of signals with lower level of noise allows to reach much higher rejection efficiency even with slower heat signals. Dependence of the rejection efficiency (by using the mean-time method) on the signal-to-noise ratio for heat signals confirms the assumption (see Fig. 5).

**Table 1.** Rejection efficiency of randomly coinciding 2ν2β events by pulse-shape discrimination of light and heat signals under condition that start of the signals is known from the generation procedure.

| Channel, rise time (ms) | Mean-time method, % | Front edge analysis, % | $\chi^2$ method, % |
|---|---|---|---|
| Light, 3 ms | 97.7±0.5 | 91.7±0.5 | 97.5±0.5 |
| Heat, 13.6 ms | 99.4±0.2 | 98.7±0.2 | 99.4±0.2 |

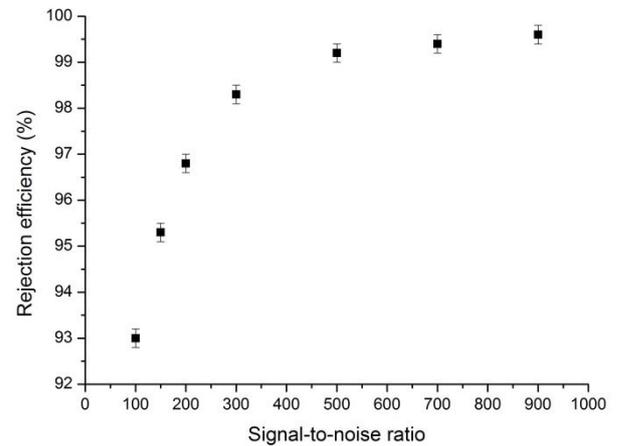

**Figure 5.** Dependence of the rejection efficiency on the signal-to-noise ratio for the heat channel. Mean-time method was used for calculation.

We tried also to analyse a dependence of the rejection efficiency on the rise time of light pulses (see Fig. 6). One could expect that for faster signal any methods should give a higher efficiency of pulse-shape discrimination. However, the trend of the rejection efficiency improvement for faster signals is rather weak. Furthermore, the rejection efficiency for the pulses with the rise time 2 ms is even worse in comparison to slower signals with the rise time 3 ms and 4.5 ms. This feature can be explained by a rather low sampling rate (1.9841 kSPS) used for the data acquisition. Indeed, the pulse profiles acquired with this sampling rate are too discrete: for instance, the front edge of the signals with the rise time 2 ms is represented by only 4 points. Such a low discretization even provides difficulties to set the acceptance factor of single events at the certain level (95% in our case), particularly in the front edge analysis.

Such an assumption was proved by taking the noise baselines data with 2 times lower sampling rate (we have transformed the data simply summarizing two channels to one). The rejection efficiency decreases in this case, as one can see in Fig. 6.

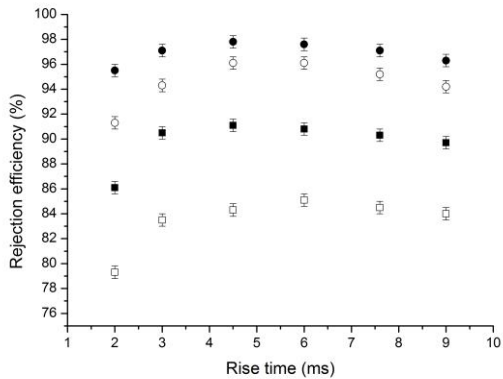

**Figure 6.** Dependence of the rejection efficiency for the light pulses (by using the mean-time method) on the rise time, signal-to-noise ratio and data acquisition sampling rate. The filled squares (circles) represent the data with the signal-to-noise ratio of 30 (100) acquired with the sampling rate 1.9841 kSPS, while the open markers show results for the same signals acquired with the sampling rate 0.9921 kSPS.

## 5 Conclusions

Random coincidence of $2\nu2\beta$ decay events could be one of the main sources of background for $0\nu2\beta$ decay in cryogenic bolometers because of their poor time resolution, particularly for $^{100}$Mo due to comparatively short half-life relatively to the two neutrino mode. However, this background can be effectively suppressed with the help of pulse-shape discrimination.

The randomly coinciding $2\nu2\beta$ decay signals were discriminated with an efficiency on the level of 99% by applying the mean-time approach to the heat signals from $ZnMoO_4$ cryogenic bolometer with the rise time about 14 ms and the signal-to-noise ratio 900, and on the level of 98% for the light signals with 3 ms rise time and signal-to-noise ratio 30 (however, the last estimation was obtained for the signals with *a priory* known start position). Other methods ($\chi^2$ approach and front edge analysis) provide comparable rejection efficiencies, however results of the front edge analysis are slightly worse due to the certain problems to distinguish events when one of the randomly coinciding signals is too small to be detected by this method. This makes the $ZnMoO_4$ cryogenic scintillating bolometers very promising to search for neutrinoless double beta decay on the level of sensitivity high enough to determine the neutrino mass hierarchy.

The signal-to-noise ratio looks the most important feature to reject randomly coinciding events, particularly in $ZnMoO_4$ due to the comparatively low light yield, which leads to a rather low signal-to-noise ratio in the light channel.

Development of algorithms to find origin of a signal with as high as possible accuracy is requested to improve rejection capability of any pulse-shape discrimination technique. Sampling rate of the data acquisition should be high enough to provide effective pulse-shape discrimination of randomly coinciding events. Finally, any pulse-shape discrimination methods should be optimized taking into account certain detector performance to reduce the background effectively.

## Acknowledgments

The work was supported in part by the project "Cryogenic detector to search for neutrinoless double beta decay of molybdenum" in the framework of the Programme "Dnipro" based on Ukraine-France Agreement on Cultural, Scientific and Technological Cooperation. The study of $ZnMoO_4$ scintillating bolometers is part of the program of ISOTTA, a project receiving funds from the ASPERA 2nd Common Call dedicated to R&D activities.